%% file: LHCP2014-Antonioli.tex
\documentclass[10pt]{article}
\usepackage{graphicx}
\usepackage{amsmath}
\usepackage{xspace}
\usepackage[square,sort,comma,numbers]{natbib}
\newcommand\pubnumber{ }

\newcommand\pubdate{\today}

\def\Title#1{\begin{center} {\Large #1 } \end{center}}
\def\Author#1{\begin{center}{ \sc #1} \end{center}}
\def\Address#1{\begin{center}{ \it #1} \end{center}}

\newcommand\pubblock{\rightline{\begin{tabular}{l} Proceedings of the Second Annual LHCP\\ \pubnumber\\
         \pubdate  \end{tabular}}}

\newenvironment{Abstract}{\begin{quotation} \begin{center} 
             \large ABSTRACT \end{center}\bigskip 
      \begin{center}\begin{large}}{\end{large}\end{center} \end{quotation}}

\newenvironment{Presented}{\begin{quotation} \begin{center} 
             PRESENTED AT\end{center}\bigskip 
      \begin{center}\begin{large}}{\end{large}\end{center} \end{quotation}}

\def\Acknowledgements{\bigskip  \bigskip \begin{center} \begin{large}
             \bf ACKNOWLEDGEMENTS \end{large}\end{center}}

\input econfmacros.tex

\newcommand{\PbPb}{\mbox{Pb--Pb}\xspace}
\newcommand{\pPb}{\mbox{p--Pb}\xspace}
\newcommand{\pp}{pp\xspace}
\newcommand{\jpsi}{J/\ensuremath{\psi}\xspace}
\newcommand{\Upsi}{\ensuremath{\Upsilon}\xspace}

\newcommand {\pT}        {\ensuremath{p_{\mathrm{T}}}\xspace}
\newcommand {\meanpT}    {\ensuremath{\langle p_{\mathrm{T}} \kern-0.1em\rangle}}
\newcommand{\KZero}     {\mbox{$\mathrm {K^0_s}$}\xspace}
\newcommand{\DZero}     {\mbox{$\mathrm {D^0}$}\xspace}
\newcommand{\DPlus}     {\mbox{$\mathrm {D^+}$}\xspace}
\newcommand{\Lam}   {\mbox{$\mathrm{\Lambda}$}\xspace}

\newcommand{\sqs}  {\ensuremath{\sqrt{s}}}
\newcommand{\seff} {\ensuremath{\sigma_{\mathrm{eff}}}\xspace}

\usepackage{color}

\def\affiliation{
INFN - sezione di Bologna, \\
Via Irnerio 46 \\
40126 Bologna BO, Italy}

\begin{document}

\large
\begin{titlepage}
\pubblock

\vfill
\Title{Double Parton Scattering, Multi-Parton Interactions, underlying event and identified hadrons: summary of recent results}
\vfill

\Author{ Pietro Antonioli  }
\Address{\affiliation}
\vfill
\begin{Abstract}

Recent results related to Double-Parton Scattering (DPS) and Multi-Parton Interactions (MPI) from the LHC experiments (ALICE, ATLAS, CMS and LHCb) are reviewed and discussed together with a brief overview of relevant literature. The robust evidence collected for DPS in different channels at LHC energies 
is complemented by an increasing understanding of our description of MPI in high energy 
collisions and the corresponding modelling of the underlying event (UE) in hadronic interactions. Potential new results expected during Run 2 at the LHC are also anticipated.

The relation and the interplay between the relevant observables for DPS, MPI and UE  analyses are discussed presenting recent attempts to bring together
their description in a single Monte Carlo tune. Identified hadron spectra at the LHC have been now measured by all collaborations and results are reviewed with an emphasis on strangeness production and baryon/meson ratio. 
The data collected during Run 1 at the LHC with different collision systems (\pp, \pPb, \PbPb) show that several particle production features appear to be more correlated with the event multiplicity than the collision system itself.

\end{Abstract}
\vfill

\begin{Presented}
The Second Annual Conference\\
 on Large Hadron Collider Physics \\
Columbia University, New York, U.S.A \\ 
June 2-7, 2014
\end{Presented}
\vfill
\end{titlepage}
\def\thefootnote{\fnsymbol{footnote}}
\setcounter{footnote}{0}
%

\normalsize 


\section{Introduction}
In this paper I briefly summarize in Section~\ref{sec:dpsmpi} the role played by Multi-Parton Interactions (MPI) in high energy hadron collisions moving then to our current description and understanding of Double Parton Scattering (DPS). In the same 
section I review also the results obtained on DPS by the different LHC experiments so far in different channels. Even if the emphasis
is on DPS results in \pp I cover also results and phenomenological predictions related to DPS and MPI in other collision
systems and with other observables like heavy flavour production as a function of the multiplicity or the study
of azimuthal correlations. The description of MPI is also key for a satisfactory modelling of the underlying event (UE) 
in the many Monte Carlo generators available. The observables and characteristics of the UE at the LHC have been studied
extensively with a wealth of papers using different approaches. Some of them are summarized and discussed in Section~\ref{sec:ue},
together with a description of the most recent studies trying to tune successfully both hard and soft scales in MPI,
respectively affecting DPS and UE observables.

Identified hadron spectra are discussed in Section~\ref{sec:idhad}. The main highlights selected here are the strangeness
and the baryon/meson ratios where Monte Carlo generators are not yet describing satisfactorily data. At the LHC identified particle
yields have been studied as a function of multiplicity and in different collision systems: I review then the
main results and the Monte Carlo tunes that seem to better describe the data.  

Given the wide range of themes presented and the space allocated both for the presentation and these proceedings, this paper 
should be intended as a selection of topics with some emphasis on more recent results, not as a comprehensive review
or a complete summary of all experimental results obtained at the LHC so far about these topics. In these proceedings I did not insert figures, giving on purpose more space to the narrative and the relevant bibliography. This paper should be therefore read alongside the 
corresponding slides~\cite{myslides} available at the conference web site. Reported results and references are updated at the time the talk 
was given. References for Monte Carlo generators and tunes are generally not reported but can be found in the referenced
corresponding experimental paper.

\section{Double Parton Scattering and Multi-Parton Interactions}\label{sec:dpsmpi}

The idealistic picture of just two partons involved in the hard scattering of a hadron-hadron collision 
with all the other spectator partons simply contributing to the underlying event is known to be 
somehow a na\"ive approximation. The possibility of having several independent parton-parton interaction in a single collision
was soon realized and already in 1987 Sj\"ostrand and Van Zijl~\cite{SjoMPI} introduced an impact parameter picture to
describe MPI. The role played by MPI is expected to increase with the collision energy: at greater center of mass energy there is
a greater access to the low-x region where the parton density increases and so, in turn, increases the MPI probability to
occurr. A devoted series of workshops is now held on this subject and their summaries~\cite{MPI2010,MPI2012} are a precious resource to review the existing literature. 

At the LHC during Run 1 many examples of analyses have been collected where Monte Carlo generators fail completely to describe data if MPI simulation
is turned off. Depending on the energy scale of the MPI a second hard scattering can occur, known 
as Double Parton Scattering (DPS). Typically the second hard scale is O(10 GeV). This opens the possibility to heavy flavour production due to MPI. At LHC energies the presence of DPS can also create a critical background to exclusive physics channels under study and it is therefore of paramount importance to achieve a satisfactory description of it.

In the usual formula for single parton scattering (SPS) the parton distribution functions (PDF) for the two involved partons and the cross section of the hard process are present. Depending on the process considered  many products can be of course produced, like four jets. The same outcome may arise from a DPS where, following Blok, Dokshitzer and others~\cite{DPSDok} double parton generalized distribution functions can be introduced. In this case, through appropriate factorization, the two hard scattering cross sections and the usual individual PDFs are re-obtained together with an additional term that depends on the transverse momentum conjugate of the relative distance between the two partons.

This factor is often called effective cross-section \seff but indeed it measures the size 
of the partonic core where the flux of short distance partons is confined. It is not really a cross section (in the sense of the 
strength of an interaction) and is therefore proportional to the transverse area of parton overlap. It reflects also 
the longitudinal parton correlations. The study of the DPS allows therefore to probe the hadronic structure and
during last years we had four papers published by LHC experiments presenting results on DPS.

%

CMS~\cite{CMSDPSwjj} and ATLAS~\cite{ATLASDPSwjj} analyses are both on W plus 2-jet events. 
These events are good candidates for this study because from one side a clean tag is 
possible for the W (using muon or electron decay) 
and  the di-jet production has a large cross section. Specific variables are identified to discriminate between as SPS, producing the same final states, and a DPS, producing respectively the W and the di-jet system. 
Pseudo-data were generated switching off DPS in the MC to create templates to then fit the distribution. Doing this exercise there is the general challenge to 
of defining a hard-scale between MPI and DPS. CMS used also the azimuthal angle between the W and the di-jet system (which is instead expected back to back in the SPS). A fit to the data using the templates allow to extract the fraction of events due to the DPS scattering
which is found to be $\approx$ 6\%. The effective cross section is finally obtained measuring also the ratio  of W events without and with two jets and the inclusive two jets cross-section.

Within uncertainties the CMS and ATLAS results are in good agreement each other and with previous measurements also, in particular
from Tevatron. A compilation of DPS results (including Tevatron, SPS and ISR) was published in~\cite{D0DPS}. It is not possible to draw a 
conclusion on the potentially weak dependence of \seff with the center of mass of energy but Run 2 might tell us more.  
Interestingly during 2014 we had also a new result from D0 at Tevatron using the channel $\gamma$ +3 jets~\cite{D0DPS} with one of the jets b/c tagged. This makes the measure sensitive to differences in the transverse spatial distribution of light and heavy quarks.  
It is good to see the advancements achieved extracting the DPS contribution using progressively more selective channels 
(four jets, $\gamma$ + 3 jets, W + 2 jets) and refining analyses techniques.

This is indeed in line with the results presented by LHCb, investigating double \jpsi events~\cite{LHCbDoubleJpsi}  and \jpsi production together with an open charm (C)~\cite{LHCbDPSjpsi,LHCbDPSjpsiERR} hadron.  Their result on the double \jpsi cross-section is found compatible with SPS contribution only. It has been pointed out that in future it will be important to make this research in channels where the DPS is expected to be the dominant production channel as the \jpsi\Upsi pair~\cite{Novoselov}. LHCb furter studied events with a \jpsi and an open charm hadron or double same-charge charm hadrons $CC$. They used then a $C\overline{C}$ events as a control sample. In these analyses the global yield has been extracted with a fit to the two-dimensional invariant mass distribution of the two particles studied, with a template for the combinatorial background. The absence of azimuthal and rapidity correlations for \jpsi~C and $CC$ events provides support for the DPS interpretation, while the presence of these correlations in the $C\overline{C}$ sample as expected
by gluon splitting is a further evidence. The values of \seff that can be obtained (combining the results for the individual
processes and for DPS, all measured by LHCb) are in agreement with the results from Tevatron, CMS
and ATLAS considering the \jpsi~C sample, while higher values (but still in reasonable agreement) are obtained via the $CC$ sample. This agreement is very interesting because LHCb, thanks to its forward geometry, is probing a different x-region. 
CMS analyzes the yield as a function of the rapidity difference between the \jpsi pair. It was indeed suggested~\cite{Maciula} that at sufficiently high values of that difference the DPS becomes the dominant production mechanisms of the pair. This could be the origin of the excess reported by CMS so far only at preliminary level~\cite{CMSdoublejpsi}.

Finally LHCb obtained also the first measurement at the LHC in the ZC channel~\cite{LHCbDPSzc} collecting a handful of events in the channels
Z\DZero and Z\DPlus, with the Z boson observed in the di-muon channel. In this case, assuming the \seff value from CDF, the measured values are within expectations for Z\DZero and below expectations for Z\DPlus. In this case, due to the probed x-region, the usual factorization ansatz used for DPS estimations could be largely violated. The next step, increasing statistics, will be to look at differential distributions for a deeper understanding of these events.

Interestingly the DPS contribution is used now also routinely to fit single mesons spectra measured by different experiments~\cite{Maciula}:
this generally helps to better describe data and this contribution is expected to become comparable with SPS at LHC Run 2 energies. As discussed by Cazaroto et al.~\cite{Cazaroto} this holds true even including parton saturation effects. The production of a bc pair will have the same cross
section that an SPS b production (that is DPS with creation of a b quark and a c quark with respect to SPS creation of a b quark).

Already at Run 1 ALICE and CMS reported measurements of c and b production as a function of multiplicity respectively for J/psi and D mesons~\cite{ALICEmultJpsi}  and for the \Upsi~\cite{CMSmultUpsi}. The relative yield shows a linear increase with the relative charged particle density for \pp, \pPb and \PbPb collisions that can be interpreted as MPI happening at a harder scale. Looking to the future several authors pointed out that such collision systems can be an important place where to study MPI and DPS during Run 2. In nuclei-nuclei collisions it is possible to have DPS involving partons belonging to different nucleons. This has the advantage to filter out longitudinal correlations~\cite{DPSinPbPb}. A sizeable fraction of DPS events in \PbPb is also expected, with - for example - the ratio of the double \jpsi production with respect to single \jpsi estimated to reach 35\% in the most central collisions~\cite{DEnt1}. The long waited smoking gun for DPS, that is double same-sign W production, could be eventually seen in \pPb collision~\cite{Dent2}. 

Interestingly ALICE is studying azimuthal correlations in \pPb also as a mean to study the role played by MPI. In 2013 new observables (the so-called uncorrelated seeds) were introduced by ALICE~\cite{ALICEazcorr} in \pp starting from azimuthal correlations between a trigger particle with a given \pT and all the others. The near and away sides are fitted with a proper combination of gaussians that allow to extract the average particle yield. Taking into account the number of trigger particles and the total number
of events it is then possible to estimate an average number of uncorrelated seeds.
A proportionality between the number of MPI and the uncorrelated seeds is indeed found for Monte Carlo based on PYTHIA. In turn the number of uncorrelated seeds was found in the data increasing with the charged multiplicity, reaching higher multiplicities when increasing the center of mass energy. The pattern is qualitatively described by tested PYTHIA tunes and underestimated by PHOJET~\cite{ALICEazcorr}. While in \pp there are indications of MPI saturation (that is at the highest multiplicities an increase of MPI is improbable), this saturation pattern is not seen in \pPb~\cite{ALICEazcorrpPb}.

\section{Underlying event studies}\label{sec:ue}

A non exhaustive list of LHC results about underlying event studies is presented in~\cite{myslides} listing 17 papers. Not all of them are discussed in the following text, in particular the papers related to the study of the forward region~\cite{LHCbue1,LHCbue2,CMSueforward,LHCbue3}. As a general comment it is interesting  to note the trend towards more sophisticated approaches and observables used for these analyses, reflecting also their complexity.

In the ''traditional'' approach the density of charged tracks, the \pT sum density and the ratio of these two quantities
are studied in spatial regions determined by a ``leading object'', being a jet or a high \pT track.
The leading object defines a ''toward'' azimuthal region and correspondingly an ``away'' and two transverse regions that are especially inspected for UE studies. The underlying event is made of particles that arise from beam-beam renmants (BBR) and the MPIs that accompany the hard scattering and they are expected to predominantly populate the transverse regions.  

The first bunch of ``traditional'' approach results at the LHC showed that pre-LHC tunes underestimated 
the measured values of the \pT sum distributions at different levels~\cite{ATLASue1}. Predictions and observables were firmly based on seminal work at Fermilab. New PYTHIA tunes as Z1 and 4C better describe data:  tune 4C describes generally better the energy dependence but Z1 better the underlying event features~\cite{CMSue1,CMSue2}.
The increase of the hadron activity with the center of mass energy increases faster in the UE~\cite{ALICEue}, consistently with predictions of models that considered harder parton scatterings to happen. Even if several tuning adjustments were necessary, we should keep in mind as a summary message that pre-LHC predictions have been very successfull: pre-LHC tune for the UE transverse charged density predicted a factor 2 between 0.9 and 7 TeV in the 'plateau' region (for leading track-jet \pT greater than 10 GeV/c)~\cite{UEreviewRF}, which is what LHC experiments then measured~\cite{CMSue1,CMSue2,ATLASue1}.


The refinement of UE analyses along years has followed two main directions: (a) the usual traditional analysis was repeated for special categories of events or (b) new observables have been introduced. 

The CMS analysis of the UE in Drell-Yan events~\cite{CMSuedy} is an example
of the former: these events are naturally much cleaner from final state radiation (FSR) contributions and using a cut on the transverse mass of the muon pair it is possible to limit the initial state radiation (ISR) contributions thus better selecting the MPI contributing to the UE activity. PYTHIA tunes based on LHC data work quite well for these events too. This tests the universality of the UE activity. The independence of the total \pT density from the invariant mass of the muon pair is then an indication of a MPI constant rate down to 40 GeV. Unsurprisingly the use of a event generator with "MPI off" fails completely to describe data.

ATLAS provided instead a comprehensive analysis~\cite{ATLASuejet} of UE activity using jet as leading object and studying then the dependence of various observables (like the particle density) as a function of the radius of the cone used to define the jet (in this sense it is representative of the second class of UE analyses). This allows to study the interplay between pQCD and soft-QCD in MC models and it is seen that PYTHIA 6 LHC tunes make a remarkable pretty job, while others require further tuning (for example 4C).

CMS elaborated then a different approach: once a jet is identified all the intrajet particles are removed and 'all the rest' is considered UE, without the traditional regions definition. The study~\cite{CMSueintra} was performed as a function of multiplicity. This is very interesting given a range of observations that points to the existence of some underlying mechanism not taken into account in event generators (like the presence of the ridges, of high multiplicity tails and further studies on event shapes and sphericity).
An increase of the hardness of the spectra as a function of multiplicity is expected for underlying and global particles (with the multiplicty increasing due to the number of semi-hard scatterings happening) while an opposite behaviour for intrajet particles is predicted. Both trends are observed. Interestingly PYTHIA tunes describes the data (with the tune Z2* performing better) but HERWIG was found not predicting that behaviour for the UE event.

Additional observables to characterize the UE were studied by CMS~\cite{CMSueneutral} looking at neutral strangeness (via \KZero and \Lam identification): the total \pT  distributions show similar trends with respect to what seen for charged particles, with the plateau again starting at \pT$\approx$10 GeV/c of the leading charged particle. This is once more consistent with an MPI activity modelling 
correlated with the centrality of the collision and it signals that hadronization and MPI are decoupled. So far we had only a limited amount of results using particle identification in UE analyses and we should expect in future these analyses able to provide further insights~\cite{UEreviewRF}. Additional classes of events studied (and special tunes elaborated~\cite{CMSuett}) are $t\overline{t}$: these events have a higher FSR activity and, in this sense, it is therefore a complementary study to the Drell-Yan mentioned before.

Very recently ATLAS published an important update of its analysis~\cite{ATLASuetrans} of UE in jet events. Besides extending the phase-space coverage the two main novelties are the detailed study of the activity in the transverse region (identifying the so-called trans-min and trans-max regions) and the computation of the UE observables separately for inclusive jet events and exclusive di-jet selections. The former approach, applied successfully before at Tevatron
but for the first time here for LHC data, helps separate BBR and MPI components of the UE and - in turn - the trans-diff
observables should be very sensitive to ISR and FSR. The latter approach revealed that the \pT density in the transverse region remains almost constant for the di-jet selection for \pT-jet greater than 20 GeV/c, while in inclusive events increases with the leading jet energy: this is once more a clear indication that pure MPI activity is independent of the hard process scale.

Finally it is interesting to report recent results that put together these very much interlinked themes in sections~\ref{sec:dpsmpi} and~\ref{sec:ue}. CMS revisited~\cite{CMSdps4j} its 4 jets result in terms of MPI/DPS pointing to the fact that a better tuning of the UE is needed before a DPS component can be extracted for this class of events. 
In a subsequent analysis note~\cite{CMStuningall} it was studied the interplay of the tuning optimised for UE or for DPS, bringing to different predictions for the \seff values that can be extracted from Monte Carlo data. Using DPS optimized tunes (via CMS data) they found then that the UE event activity as measured by ATLAS would be underpredicted. This is an indication of some tension in the simultaneous description of soft and harder MPI, at least in these tested PYTHIA tunes. A more successfull attempt was instead reported for a HERWIG++ tune by~\cite{HERWIGtune}.

\section{Identified hadrons}\label{sec:idhad}
Also for identified hadrons in \pp collisions we had a first bunch of results from all LHC experiments that brought very useful information
for tuning MC generators at LHC energies testing in particular our modelling of the hadronization process. I will review more recent results with respect to two issues still unresolved in terms of a satisfactory description of data: strangeness production and baryon/meson ratios. Despite the wide range of results comparing data and MC tunes an understanding of the underlying mechanisms doesn't seem reached. From early LHC results (for example from ALICE, ATLAS and LHCb~\cite{CMSspe1,ALICEspe1,LHCbspe1}) we know that pre-LHC tunes like D6T underpredict strange hadrons: PYTHIA6 or PYTHIA8 tunes based on LHC data as Z2 and 4C partially address this problem as reported by CMS in~\cite{CMSspe2} in terms of $K/\pi$ at low \pT or by ATLAS measuring the \KZero spectra~\cite{ATLASkzero}, but however they were found to fail up to a factor four with 
multi-strange baryons by ALICE~\cite{ALICEmultis}. It was noted by Ortiz et al.~\cite{Ortiz} that the $p/\pi$ ratio measured
by ALICE can be satisfactorily described by PYTHIA 4C (which includes color reconnection) up to \pT=4 GeV/c, but the same mechanism
helps less at rapidities measured by LHCb~\cite{LHCbspe2}, especially at very low \pT. 

ATLAS measured recently ~\cite{ATLASphi} $\phi$ production (that within the overlapping kinematical region is in agreement 
with the previous ALICE measurement~\cite{ALICEphi}) reporting comparisons with many generators: EPOS LHC (which has
some 'collective' hadronization added via a flow parameterization) better describes data but surprisingly most recent PYTHIA 6 or 8 tunes are
disfavoured with respect to PYTHIA 6 DW (which is a pre-LHC tune optimized on CDF data). ALICE tested also the ratio $\Omega/\phi$:
in hadronization models via string fragmentation this ratio is predicted being particularly sensitive to the tension of the string.
While using HIJING/BB helped to describe data in \pp collision~\cite{ALICEphi}, this is not the case for \PbPb collisions~\cite{ALICEomphiPbPb}.

During Run 1 extending the study of identified particle production features to \pPb and \PbPb collision was very instructive. Studying the average \pT of charged particle as a function of charged particles in the event it was shown by ALICE that \pp data can be described by PYTHIA when using color reconnection~\cite{ALICEmult}. For \pPb EPOS describes data where others (DPMJET, HIJING, AMPT) fail. CMS extended these results for identified spectra in different rapidity and momentum ranges~\cite{CMSmult}. Comparing \pp and \pPb collisions it also showed that particle production characteristics (both the average \pT and yield ratios) seem very much correlated to the multiplicity (more than the collision system and the center-of-mass energy). ALICE studied also the evolution of particle ratios (both $p/\pi$ and \Lam/\KZero) at fixed \pT bins: the values differ but they show the same scaling with $dN_{ch}/d\eta$~\cite{ALICEratios}. 





\section{Conclusions}
The intense analysis of LHC Run 1 data brought a substantial confirmation of our understanding and an increased comprehension of the role played by MPI in hadron-hadron collision. This is seen through the lenses of different themes, observables and analysis approaches. Processes directly sensitive to DPS have now been measured at the LHC. While the contribution arising from DPS generally brings predictions closer to measurements,   more observations are expected in Run 2. MPI are also studied via azimuthal correlations in different collision systems. After a first bunch of results (somehow repeating the successfull studies and methodologies developed at Tevatron), specific UE measurements in different categories of events are emerging as the next tool to further constrain models and tunes. Attempts to obtain Monte Carlo tunes simultaneoously
describing MPI, DPS and UE all together are also on-going.

In terms of identified spectra the strangeness abundance and the baryon/meson ratio remain areas where we need to improve predictions and our understanding of the underlying mechanisms. We have also a pattern of indications pointing to a better data agreement with MC predictions when some ``collectivity'' in the form - for example - of color reconnection or hydrodynamics is added to the generators. Intriguingly several particle production features appear more correlated with the event multiplicity than the collision system.

The study of \pPb and \PbPb collisions at the LHC could be much more revealing than originally anticipated for the topics
discussed in this talk. The collection and analysis of LHC data at \sqs=14 TeV in \pp collision will be however a first excellent
test opportunity of what we learned so far and of the predictivity of the tunes developed in the first 
five years of the LHC era.

\vspace{-0.5cm}
\Acknowledgements
\vspace{-0.2cm}
I am grateful to C.~Loizides, A.~Morsch, J.F.~Grosse-Oethringhaus, J.~Hollar, G.~Veres, K.~Mueller, V.~Belyaev, D.~Volyanskyy and S.~Glazov from the ALICE, ATLAS, CMS and LHCb Collaborations for useful input.

\end{document}

%% file: econfmacros.tex
\def\beq{\begin{equation}}
\def\eeq#1{\label{#1}\end{equation}}
\def\eeqn{\end{equation}}

\def\beqa{\begin{eqnarray}}
\def\eeqa#1{\label{#1}\end{eqnarray}}
\def\eeqan{\end{eqnarray}}

\let\bar=\overbar

\def\Dslash{\not{\hbox{\kern-4pt $D$}}}
\def\dslash{\not{\hbox{\kern-2pt $\del$}}}

\def\msb{{\bar{\ssstyle M \kern -1pt S}}}

